\documentclass{article}

\usepackage{latexsym}

\usepackage{amssymb, graphicx}
 \thinmuskip =
0.5\thinmuskip \medmuskip = 0.5\medmuskip \thickmuskip =
0.5\thickmuskip \arraycolsep = 0.3\arraycolsep

\newcommand{\R}{\mathbb{R}}
\newcommand{\fv}{f_{\wedge}}
\newcommand{\Ev}{E_{\wedge}}
\newcommand{\Bv}{B_{\wedge}}
\newcommand{\Evin}{E_{\wedge}^\mathrm{in}}
\newcommand{\Bvin}{B_{\wedge}^\mathrm{in}}
\newcommand{\rhov}{\rho_{\wedge}}
\newcommand{\jv}{j_{\wedge}}
\newcommand{\fvin}{\fv^{\mathrm{in}}}

\newcommand{\pv}{\mathfrak{p}_\wedge}

\def\prfe{\hspace*{\fill} $\Box$

\smallskip \noindent}
\newtheorem{Theorem}{Theorem}
\newtheorem{Proposition}{Proposition}
\newtheorem{Lemma}{Lemma}
\newtheorem{Definition}{Definition}
\newtheorem{Corollary}{Corollary}

\title{On a characteristic initial value \\problem in plasma physics}
\author{Simone Calogero \\
Institutt for matematiske fag\\
NTNU Alfred Getz' vei 1 N-7491\\
Trondheim, Norway\\
E-mail: simone.calogero@math.ntnu.no}
\date { }
\begin{document}
\maketitle

\begin{abstract}
The relativistic Vlasov-Maxwell system of plasma physics is
considered with initial data on a past light cone. This
characteristic initial value problem arises in a natural way as a
mathematical framework to study the existence of solutions
isolated from incoming radiation. Various consequences of the
mass-energy conservation and of the absence of incoming radiation
condition are first derived assuming the existence of global smooth solutions.
In the spherically symmetric case, the existence of a
unique classical solution in the future of the initial cone
follows by arguments similar to the case of initial data at time
$t=0$. The total mass-energy of spherically symmetric solutions
equals the (properly defined) mass-energy on backward and forward
light cones.
\end{abstract}

\section{Introduction}\label{intro}
\setcounter{equation}{0} In a system of
Cartesian coordinates $(t,x)$, $t\in\R,\,x\in\R^3$, the
Vlasov-Maxwell system is given by
\begin{equation} \label{vlasovt}
\partial_t f +\widehat{p}\cdot\nabla_x f +(E+\widehat{p}\times B)\cdot\nabla_p f = 0,
\end{equation}
\begin{equation}\label{maxwellt}
\partial_t E-\nabla\times B=- j,\quad\partial_tB+\nabla\times E=0,
\end{equation}
\begin{equation}\label{constrainteqt}
\nabla\cdot E=\rho,\quad\nabla\cdot B=0,
\end{equation}
\begin{equation} \label{sourcedeft}
\rho(t,x) = \int f(t,x,p)\,dp,\quad j(t,x)=\int \widehat{p}\,f(t,x,p)\,dp.
\end{equation}
The Vlasov-Maxwell system models the dynamics of collisionless
plasmas. We consider for simplicity a plasma consisting of a
single species of particle. The unknowns are the particle density
in phase-space, $f=f(t,x,p)$, where $p \in \R^3$ is the momentum
variable, and the mean electromagnetic field $(E,B)=(E,B)(t,x)$
generated by the particles. The expression
\[
\widehat{p} =\frac{p}{\sqrt{1+|p|^2}}
\]
denotes the relativistic velocity of a particle with momentum $p$.
Units are chosen such that the mass and the charge of each
particle and the speed of light are equal to unity. The simbol
$\times$ denotes the usual vector product in $\R^3$. We refer to
\cite{BGP, C2, DipLio, GS, GS2, GSh2, H, KS, R} for background on
the Cauchy problem for the Vlasov-Maxwell system. Classical
solutions of (\ref{vlasovt})--(\ref{sourcedeft}) satisfy the
energy identity
\begin{equation}\label{localcon}
\partial_te + \nabla\cdot\mathfrak{p}=0,
\end{equation}
where
\[
e(t,x) = \int \sqrt{1+|p|^2}\,f\,dp + \frac{1}{2}|E|^2 +
\frac{1}{2}|B|^2, \quad \mathfrak{p}(t,x) = \int p\,f\,dp +E\times
B.
\]
Integrating (\ref{localcon}) one obtains the conservation of the
total energy
\begin{equation}\label{totalcons}
M(t)=\int\int\sqrt{1+|p|^2}
f\,dp\,dx+\frac{1}{2}\int\left(|E|^2+|B|^2\right)\,dx=const.
\end{equation}
Solutions of Vlasov-Maxwell also satisfy the continuity
equation
\begin{equation}\label{conteq}
\partial_t\rho + \nabla\cdot j=0,
\end{equation}
which upon integration leads to the conservation of the total (rest) mass
\begin{equation}\label{totalmasscons}
N(t)=\int\int f\,dp\,dx=const.
\end{equation}
The purpose of this paper is to set up a mathematical framework
for the analysis of solutions to the Vlasov-Maxwell
system which satisfy the no-incoming radiation condition, that is
\begin{equation}\label{nirc}
\lim_{r\to\infty}\int_{v_1}^{v_2}\int_{|x|=r}k\cdot [E\times
B](v-|x|,x)\,dS_r(x)\,dv=0,
\end{equation}
for all $v_1,v_2\in\R$, where $dS_r$ is the surface element on the
sphere of radius $r$ and $k=x/r$ is the unit normal on this
sphere. This corresponds to the physical condition that the
electromagnetic field carries no energy to the past null infinity
of Minkowski space, see \cite{C2,C3}.

Solutions of Vlasov-Maxwell isolated from incoming radiation were
first studied in \cite{C2}. The result of \cite{C2} is that such
solutions exist globally in time for small data of the Cauchy
problem (i.e., data at time $t=0$). However the Cauchy problem is
not a natural framework to generate solutions isolated from
incoming radiation. In fact, since the no-incoming radiation
condition is imposed at $t\to -\infty$, there is no meaningful
notion of local isolated solution with data at $t=0$. Therefore,
in the framework of the Cauchy problem, one can only prove the
existence of global (or semiglobal) solutions which satisfy
(\ref{nirc}). This requires the use of uniform in time \textit{a
priori} estimates, which are not available in general for
non-linear problems.

A more natural setting for the study of isolated solutions is the
initial value problem with data on a surface which cuts past null
infinity. Examples of such surfaces are past light cones and
backward hyperboloids. This paper is concerned with the first
case.

Another motivation for studying the initial value problem with
data on a past light cone comes from physical grounds.  The
initial data correspond to the outcome of an experimental
measurement on the state of the physical system at the present
time; the existence of a unique solution with the given data
assures that the outcome of any future measurement is predicted by
the theory. If this physical interpretation of the initial value
problem is accepted, then it is clear that the initial data for
relativistic models, such as the Vlasov-Maxwell system, should be
given on a past light cone. In fact the set of events which are
accessible to an observer at the proper time $t=0$ lie on the past
light cone with vertex on the world line of the observer at $t=0$.
The state of the system on the surface $t=0$, on the other hand,
cannot be measured, because these events form a spacelike
hypersurface in Minkowski space. Such a discrepancy between the
``physical'' and the ``mathematical'' initial value problem has
been sometimes discussed in the physical literature, see
\cite{Da,De,E} and the references therein.

In order to study the Vlasov-Maxwell system with initial data on a
past light cone, we first rewrite the equations in the coordinates
$(v,x)$, where $x\in\R^3$ and $v\in\R$ is the advanced time, which
is defined by the condition that the surfaces $v=$constant
correspond to the past light cones with vertex on the timelike
curve $|x|=0$ (the world-line of the observer). Denote by
$\fv=\fv(v,x,p),\,(\Ev,\Bv)=(\Ev,\Bv)(v,x)$ the particle density
and the electromagnetic field expressed in these coordinates. They
are related to the solutions of
(\ref{vlasovt})--(\ref{sourcedeft}) by
$\fv(v,x,p)=f(v-|x|,x,p),\,(\Ev,\Bv)(v,x)=(E,B)(v-|x|,x)$ and
therefore they satisfy the equations
\begin{equation}\label{vlasov}
(1+\widehat{p}\cdot
k)\partial_v\fv+\widehat{p}\cdot\nabla_x\fv+(\Ev+\widehat{p}\times\Bv)\cdot\nabla_p\fv=0,
\end{equation}
\begin{eqnarray}
&&\partial_v(\Ev-k\times\Bv)=\nabla\times\Bv-\jv,\label{maxwell1}\\
&&\partial_v(\Bv+k\times\Ev)=-\nabla\times\Ev,\label{maxwell2}
\end{eqnarray}
\begin{eqnarray}
&&\partial_v(\Ev\cdot k)+\nabla\cdot\Ev=\rhov,\label{constrainteq1}\\
&&\partial_v(\Bv\cdot k)+\nabla\cdot\Bv=0,\label{constrainteq2}
\end{eqnarray}
where
\begin{equation}\label{sourcedef}
k=\frac{x}{|x|},\quad \rhov(v,x)=\int
\fv\,dp,\quad\jv(v,x)=\int\widehat{p}\,\fv\,dp.
\end{equation}
Initial data are given at $v=0$ and denoted by
\[
\fvin(x,p)=\fv(0,x,p),\quad\Evin(x)=\Ev(0,x),\quad \Bvin(x)=\Bv(0,x).
\]
Later we shall discuss the equivalence of the system above
with the evolution equations (\ref{vlasov})-(\ref{maxwell2}) and a set of constraint equations on the initial data.

In this paper we are interested in the question of existence and
uniqueness of classical solutions in the future (i.e., for $v\in
[0,\infty[$)  which match the initial data at $v=0$. Obviously,
one cannot expect (in general) that a unique solution is
determined by initial data at $v=0$, since the intersection
between the initial surface and the domain of dependence of the
solutions on a space-time point in the future is not a compact
set. However it turns out that the Maxwell equations
(\ref{maxwell1})--(\ref{constrainteq2}) have indeed at most one
solution for given data $(\Evin,\Bvin)$ at $v=0$ provided the
no-incoming radiation condition is satisfied. This suggests that
the solutions we seek to the initial value problem with data on a
past light cone should be restricted to the class of solutions
isolated from incoming radiation.

This paper is organized as follows. In Section \ref{ivp} we prove
some general properties of smooth solutions to the system
(\ref{vlasov})--(\ref{sourcedef}). The results of Section
\ref{ivp} are conditional, as they assume the existence of global
classical solutions. In Section \ref{ivp} we also discuss the
relation between the conservation laws satisfied by solutions with
data on a past light cone and solutions with data at $t=0$. Note
in fact that for solutions with data on a past light cone, the
conservation of the total mass and of the total energy are not
obvious. In Section \ref{sphsymsol} we prove global existence and
uniqueness of spherically symmetric solutions. This result is
obtained by adapting to our case the proof of global existence for
the Cauchy problem given in \cite{Ba, GSh}.  In spherical symmetry
the magnetic field vanishes identically (if decay at infinity is
imposed) and the Maxwell equations reduce to the Poisson equation
for the electric field. Hence there is neither incoming nor
outgoing radiation in spherical symmetry. We will show that, as a
consequence of the absence of radiation, spherically symmetric
solutions satisfy the conservation laws (\ref{totalcons}),
(\ref{totalmasscons}) and that the total mass-energy equals the
mass-energy on the past light cones and on the future light cones.

In a subsequent publication the results of this paper will be
extended to the Nordstr\"om-Vlasov system (see \cite{C1} for a
derivation of this model). While it is easy to generalize the
formal analysis of Section 2 below to the Nordstr\"om-Vlasov
system, the proof of global existence and uniqueness of
spherically symmetric solutions is different and considerably more
involved since the Nordstr\"om scalar field equation remains
hyperbolic---and so radiation propagates---also in spherical
symmetry.

\section{The initial value problem with data on a past light
cone}\label{ivp}
\setcounter{equation}{0} An assumption on the
initial data which will be made throughout is that
\[
0\leqslant\fvin\in C^1_c(\R^3\times\R^3),\quad\Evin,\Bvin\in
C^2(\R^3)
\]
and we define
\[
R_0=\inf\{R:\fvin(x,p)=0,\,|x|\geqslant R_,\,p\in\R^3\}.
\]
Hence $\fvin=0$, for $|x|\geqslant R_0$. In this section we study
several properties of global solutions satisfying the regularity
condition
\[
\fv\in C^1([0,\infty[\times\R^3\times\R^3),\quad\Ev,\Bv\in
C^1([0,\infty[\times\R^3)
\]
and so they are solutions of (\ref{vlasov})--(\ref{sourcedef}) in
a classical sense. We also assume that $\fv$ has bounded support
in the momentum, precisely
\[
\mathcal{P}_\wedge(v)=\sup\{|p|:\fv(s,x,p)\neq 0,\,0\leqslant
s\leqslant v,\,x\in\R^3\}<\infty,\quad\forall\,v\in\R.
\]
In particular, all the integrals in the momentum variable in the
sequel are understood to be extended over a compact set. We split
the analysis in two different subsections.
\subsection{The Vlasov equation}
We start by pointing out some basic properties of $\fv$. Note
the estimate
\begin{eqnarray}
1+\widehat{p}\cdot k&=&1+\frac{p\cdot
k}{\sqrt{1+|p|^2}}\geqslant 1-\frac{|p|}{\sqrt{1+|p|^2}}\nonumber\\
&=&\frac{1}{\sqrt{1+|p|^2}(\sqrt{1+|p|^2}+|p|)}\geqslant
\frac{1/2}{1+|p|^2};\label{est}
\end{eqnarray}
hence when the support in $p$ of $\fv$ is bounded, the equation
(\ref{vlasov}) is equivalent to
\begin{equation}
\partial_v\fv+\frac{p}{p_0}\cdot\nabla_x\fv+\frac{1}{p_0}\left(\sqrt{1+|p|^2}\Ev+p\times\Bv\right)\cdot\nabla_p\fv=0,\label{vlasov2}
\end{equation}
where $p_0$ is defined by
\[
p_0=\sqrt{1+|p|^2}+p\cdot k > 0.
\]
The characteristics of the differential operator in the left hand
side of (\ref{vlasov2}) are the solutions of
\begin{equation}\label{char}
\dot{x}=\frac{p}{p_0},\quad
\dot{p}=\frac{1}{p_0}\left(\sqrt{1+|p|^2}\Ev+p\times\Bv\right)
\end{equation}
and we denote by $(X,P)(s,v,x,p)$, or simply $(X,P)(s)$, the
characteristic satisfying $(X,P)(v)=(x,p)$. Since the particle
density $\fv$ is constant along these curves, we obtain the
following representation formula for the solution of the Vlasov
equation:
\begin{equation}\label{reprf}
\fv(v,x,p)=\fvin((X,P)(0,v,x,p)).
\end{equation}
In particular $\fv$ remains non-negative for all times and
$\|\fv(v)\|_\infty\leqslant\|\fvin\|_\infty$. In the next lemma we
estimate the $x-$support of $\fv$.
\begin{Lemma}\label{supportf}
For all $v\geqslant 0$,
\[
\fv(v,x,p)=0,\quad\textnormal{{\it for} } |x|\geqslant
R_0+\frac{1}{2}v.
\]
\end{Lemma}
\noindent\textit{Proof: } For all $0\leqslant s\leqslant v$ we have, by the
first equation in (\ref{char}),
\[
|x|=|X(0)|+\int_0^v\frac{P(\tau)\cdot
K(\tau)}{\sqrt{1+|P(\tau)|^2}+P(\tau)\cdot K(\tau)}d\tau,
\]
where $K=X/|X|$. Let $[0,v]=\mathcal{I}^-\cup\mathcal{I}^+$, where
\[
\mathcal{I}^-=\{\tau\in [0,v]:(P\cdot K)(\tau)\leqslant 0\},\quad
\mathcal{I}^+=\{\tau\in [0,v]:(P\cdot K)(\tau)> 0\}.
\]
Thus, using $\sqrt{1+|p|^2}>p\cdot k$,
\begin{eqnarray*}
|x|&\leqslant& |X(0)|+\int_{\mathcal{I}^+}\frac{P\cdot
K}{\sqrt{1+|P|^2}+P\cdot K}d\tau\\
&\leqslant&|X(0)|+\frac{1}{2}\textnormal{meas}(\mathcal{I}^+)\leqslant
|X(0)|+\frac{1}{2}v.
\end{eqnarray*}
Since $|X(0)|\leqslant R_0$ in the
support of $\fv$, the lemma is proved. \prfe

We shall now derive the conservation laws satisfied by the
solutions of (\ref{vlasov2}). A straightforward computation
reveals that the right hand side of the system (\ref{char}), i.e.,
the vector
\[
F(v,x,p)=\left[\frac{p}{p_0},\frac{1}{p_0}\left(\sqrt{1+|p|^2}\Ev+p\times\Bv\right)\right],
\]
satisfies
\begin{equation}\label{identity}
\left[\nabla_{(x,p)}\cdot F\right]\left(s,X(s),P(s)\right)
=-\frac{d}{ds}\log\left(1+\widehat{P}(s)\cdot K(s)\right),
\end{equation}
where $\widehat{P}=P/\sqrt{1+P^2}$. In fact, each side of
(\ref{identity}) equals, along characteristics,
\[
-\frac{1}{(1+\widehat{p}\cdot k)^2}\left [\frac{|\widehat{p}\times
k|^2}{|x|}+\frac{1}{\sqrt{1+|p|^2}}\Big(\Ev\cdot
(k-(\widehat{p}\cdot k)\widehat{p})-(\widehat{p}\times
k)\cdot\Bv\Big)\right].
\]
From (\ref{identity}) we deduce
\[
\det\left[\frac{\partial(X,P)(s)}{\partial(x,p)}\right]=\frac{1+\widehat{p}\cdot
k}{1+\widehat{P}(s)\cdot K(s)}.
\]
Hence using (\ref{reprf}) the next lemma follows.
\begin{Lemma}
For any measurable function $Q:\R\to\R$,
\[
\int\int Q(\fv)(1+\widehat{p}\cdot k)\,dp\,dx=const.
\]
In particular, by choosing $Q(z)=z^q$, $q\geqslant 1$,
\begin{equation}\label{lqest}
\|(1+\widehat{p}\cdot
k)^{1/q}\fv(v)\|_{L^q(\R^3\times\R^3)}=const.
\end{equation}
\end{Lemma}
The case $q=1$ in (\ref{lqest}) corresponds to the conservation of
the (rest) mass on the past light cones. To be more precise, observe that $\rhov,\,\jv$ defined in (\ref{sourcedef})
satisfy the equation
\begin{equation}\label{localconsingmass}
\partial_v(\rhov+j_\wedge\cdot k)=-\nabla\cdot j_\wedge.
\end{equation}
The latter can be proved either by using (\ref{vlasov}) or by a
simple change of variable in (\ref{conteq}). We define the mass
$N_\wedge(v)$ on the past light cone at time $v$ as
\[
N_\wedge(v)=\lim_{r\to\infty}\mathfrak{n}_\wedge(v,r),\quad\mathfrak{n}_\wedge(v,r)=\int_{|x|\leqslant
r}(\rhov+j_\wedge\cdot k)\,dx.
\]
Note that the function $\mathfrak{n}_\wedge(v,\cdot)$ is
non-decreasing and so the above limits exists. By (\ref{lqest}),
$N_\wedge(v)=N_\wedge(0)$, for all $v\geqslant 0$. The total mass
of a solution, given by (\ref{totalmasscons}), can be rewritten as
\[
N(t)=\lim_{r\to\infty}\mathfrak{n}(t,r),\quad
\mathfrak{n}(t,r)=\int_{|x|\leqslant r}\rho(t,x)\,dx=\int_{|x|\leqslant r}\rhov(t+|x|,x)\,dx,\quad
t\geqslant 0.
\]
In the next lemma we prove a formula which relates the mass
functions $N(v)$ and $N_\wedge(v)$.
\begin{Lemma}\label{massidentity}
For all $v\geqslant 0$,
\[
\mathfrak{n}(v,r)=\mathfrak{n}_\wedge(v,r)-\int_v^{v+r}\int_{|x|=r}\jv\cdot
k(v',x)\,dS_r(x)\,dv'.
\]
\end{Lemma}
\noindent\textit{Proof: }Integrating (\ref{localconsingmass})
between $v$ and $v+|x|$ we get
\[
(\rhov+j_\wedge\cdot k)(v+|x|,x)-(\rhov+j_\wedge\cdot
k)(v,x)=-\int_{v}^{v+|x|}\nabla\cdot j_\wedge(v',x)\,dv'.
\]
Integrating in the region $|x|\leqslant r$ we get
\begin{eqnarray}\label{coneint3}
\int_{|x|\leqslant r}(\rhov+j_\wedge\cdot k)(v+|x|,x)\, dx
&=&\int_{|x|\leqslant r}(\rhov+\jv\cdot k)(v,x)\, dx\\
&&-\int_{|x|\leqslant r}\int_{v}^{v+|x|}\nabla\cdot
\jv(v',x)\,dv'\,dx.\nonumber
\end{eqnarray}
Now we use the identity
\[
\nabla\cdot \int_{v}^{v+|x|}\jv(v',x)\,dv'= (\jv\cdot
k)\,(v+|x|,x) +\int_{v}^{v+|x|}\nabla\cdot \jv(v',x)\,dv'.
\]
Substituting into (\ref{coneint3}) and using the Gauss theorem
proves the lemma.\prfe

By Lemma \ref{supportf} all characteristics of the Vlasov equation
must cross the surfaces $t=v+|x|=const.$ for all $t\geqslant 0$ in
compact sets of $x$. This means in particular that no mass can be
lost at spacelike infinity, which explains why the following lemma
holds true.
\begin{Lemma}\label{Nwedge=N}
For all $v\geqslant 0$, $N(v)=N_\wedge(v)=N_\wedge(0)$.
\end{Lemma}
\noindent\textit{Proof: }By Lemma \ref{supportf} and Lemma \ref{massidentity} we have
$\mathfrak{n}(v,r)=\mathfrak{n}_\wedge(v,r)$, for
$r>2R_0+v$. The claim follows by letting $r\to\infty$.\prfe

Next we define the mass on the future light cone at time $v$ as
\[
N^\vee(v)=\lim_{r\to\infty}\mathfrak{n}^\vee(v,r),\quad\mathfrak{n}^\vee(v,r)=\int_{|x|\leqslant
r}(\rho^\vee-j^\vee\cdot k)\,dx,
\]
where
\begin{eqnarray*}
&&\rho^\vee(v,x)=\rhov(v+2|x|,x)=\rho(v+|x|,x),\\
&&j^\vee(v,x)=\jv(v+2|x|,x)=j(v+|x|,x).
\end{eqnarray*}
By a change of variable in (\ref{localconsingmass}) (or in (\ref{conteq})) we have
\begin{equation}\label{localconsoutmass}
\partial_v(\rho^\vee-j^\vee\cdot k)=-\nabla\cdot j^\vee,
\end{equation}
Integrating (\ref{localconsingmass}) in time between $v$ and
$v+2|x|$ and proceeding as in the proof of Lemma
\ref{massidentity} we obtain
\begin{equation}\label{massidentity2}
\mathfrak{n}^\vee(v,r)=\mathfrak{n}_\wedge(v,r)-\int_v^{v+2r}\int_{|x|=r}\jv\cdot
k(v',x)\,dS_r(x)\,dv'.
\end{equation}
Moreover, by (\ref{localconsoutmass}), $\mathfrak{n}^\vee$
satisfies the equations
\begin{equation}\label{eqsn}
\partial_v\mathfrak{n}^\vee=-\int_{|x|=r} j^\vee\cdot k\,dS_r(x),\quad \partial_v\mathfrak{n}^\vee-\partial_r\mathfrak{n}^\vee=
-\int_{|x|=r}\rho^\vee\,dS_r(x).
\end{equation}
The evolution of the mass on the future light cones is studied in
the following lemma.
\begin{Lemma}\label{Nveedecreasing}
The function $N^\vee(v)$ is non-increasing, that is,
\[
N^\vee(v_2)\leqslant N^\vee(v_1),\ \forall\,\, v_1\leqslant v_2.
\]
Moreover
\[
(i)\quad N^\vee(v_2)=N^\vee(v_1) \textnormal{\textit{ iff }} \lim_{r\to\infty}\int_{v_1}^{v_2}\int_{|x|=r} j^\vee\cdot k\,\,dS_r(x)\,dv=0;
\]
\[
(ii)\quad N^\vee(v)=N_\wedge (v) \textnormal{\textit{ iff }} \lim_{r\to\infty}\int_v^{v+2r}\int_{|x|=r}\jv\cdot
k\,\,dS_r(x)\,dv'=0.
\]
\end{Lemma}
\noindent\textit{Proof: }Integrating the second equation in
(\ref{eqsn}) along characteristics we have
$\mathfrak{n}^\vee(v_2,r-v_2)\leqslant\mathfrak{n}^\vee(v_1,r-v_1)$,
for all $v_2\geqslant v_1$ and $r > v_2$. In the limit
$r\to\infty$ this implies that $N^\vee$ is non-increasing. The
claim (i) follows by integrating in time the first equation in
(\ref{eqsn}) on the interval $[v_1,v_2]$ and letting $r\to\infty$,
while (ii) follows by (\ref{massidentity2}), again in the limit
$r\to\infty$.\prfe

We remark that for solutions with data on a past light cone it is
not obvious that $N^\vee(v)$ is bounded. Moreover, even if
bounded, it needs not to be constant. If $N^\vee(v_2)<
N^\vee(v_1)$,  for $0\leqslant v_1< v_2$, the difference
$N^\vee(v_1)-N^\vee(v_2)$ measures the mass lost at future null
infinity in the interval $[v_1,v_2]$ of the advanced time.
Finally, even if $N^\vee$ is bounded and constant it is not
obvious that it must equal $N_\wedge$, since the limit condition
in (ii) of Lemma \ref{Nveedecreasing} might not be satisfied.

In the next lemma we show that a sufficient condition for the
limits in (i) and (ii) of Lemma \ref{Nveedecreasing} to be zero is
that the momentum support of $\fv$ is bounded {\it uniformly} in
$v\in\R$, as this condition implies that no particles can reach
future null infinity.
\begin{Lemma}\label{zerolimits}
Assume $\mathcal{P}_\wedge(v)\leqslant D$, for all $v\geqslant 0$
and for some positive constant $D$. Then, for all
$v_1,\,v_2,\,v\geqslant 0$,
\[
N^\vee(v_2)=N^\vee (v_1),\quad N^\vee(v)=N_\wedge (v).
\]
In particular, by Lemma \ref{Nwedge=N}, $N^\vee(v)=N(v)=N_\wedge
(v)=N_\wedge(0)$, for all $v\geqslant 0$.
\end{Lemma}
\noindent\textit{Proof: }By the assumption,
\[
\sqrt{1+|p|^2}\geqslant\frac{\sqrt{1+D^2}}{D}|p|\geqslant
\frac{\sqrt{1+D^2}}{D} (p\cdot k)
\]
in the support of $\fv$ and so, as in the proof of Lemma
\ref{supportf},
\[
|x|\leqslant |X(0)|+\int_{\mathcal{I}^+}\frac{P(\tau)\cdot
K(\tau)}{\sqrt{1+|P(\tau)|^2}+P(\tau)\cdot K(\tau)}d\tau\leqslant
R_0+\frac{D}{D+\sqrt{1+D^2}}\,v,
\]
for all $(x,p)\in {\rm supp}\,\fv(v)$, where
$\mathcal{I}^+=\{\tau\in [0,v]:(P\cdot K)(\tau)> 0\}$. This
implies that  $\fv(v,x,p)=0$, for $|x|\geqslant R_0+av$ and $a\in
[0,\frac{1}{2}[$. Hence
\[
\int_{v_1}^{v_2}\int_{|x|=r} j^\vee\cdot k\,dS_r(x)\,dv=0,
\textnormal{ for }r>(1-2a)^{-1}(R_0+av_2),
\]
\[
\int_v^{v+2r}\int_{|x|=r}\jv\cdot k\,dS_r(x)\,dv'=0,
\textnormal{ for }r>(1-2a)^{-1}(R_0+av).
\]
Lemma 4 concludes the proof.\prfe
\subsection{The Maxwell equations}
We now pass to study some general properties of the
electromagnetic field $(\Ev,\Bv)$. First we show the equivalence
of the Vlasov-Maxwell system
with a set of evolution equations and a set of constraint
equations on the initial data. An important point is that, in the
present situation, there are more constraint equations than in the
case of the Cauchy problem, since the initial data are given on a
characteristic surface. Computing the vector product of (\ref{maxwell1}) with the unit vector $k$, subtracting (\ref{maxwell2}) and then using (\ref{constrainteq2}) we obtain
\begin{equation}\label{temp1}
k\times\left(\nabla\times\Bv\right)-k\left(\nabla\cdot\Bv\right)+\nabla\times\Ev-k\times\jv=0.
\end{equation}
Moreover, computing the vector product of (\ref{maxwell2}) with $k$, adding (\ref{maxwell1}) and then using (\ref{constrainteq1}) we obtain
\begin{equation}\label{temp2}
\nabla\times\Bv+k\left(\nabla\cdot E_\wedge\right)-k\times\left(\nabla\times\Ev\right)-\rhov k-\jv=0.
\end{equation}
On the other hand, the equation (\ref{constrainteq1}) follows from
(\ref{maxwell1}) and (\ref{temp2}), whereas (\ref{constrainteq2})
follows from (\ref{maxwell2}) and (\ref{temp1}). Hence the whole
set of the Maxwell equations is equivalent to the system composed
by (\ref{maxwell1})-(\ref{maxwell2}) and
(\ref{temp1})-(\ref{temp2}). Clearly, since
(\ref{temp1})-(\ref{temp2}) are valid for all times, then they
must be imposed at $v=0$ in order to obtain a solution of the
initial value problem, i.e., (\ref{temp1})-(\ref{temp2}) are
constraint equations on the initial data.  However these
constraint equations are not totally independent. Let $W_1$, $W_2$
denote the left hand side of (\ref{temp1}) and (\ref{temp2}),
respectively. It is easy to verify the following identities:
\[
W_1=k\times
W_2+k\left(k\cdot\nabla\times\Ev-\nabla\cdot\Bv\right),
\]
\[
W_2=-k\times
W_1+k\left(k\cdot\nabla\times\Bv+\nabla\cdot\Ev-\rhov-\jv\cdot
k\right).
\]
From this it follows that (\ref{temp1})-(\ref{temp2}) are
equivalent to the equations
\begin{equation}\label{scalarconst}
\nabla\cdot\Bv=k\cdot\nabla\times\Ev,\quad
k\cdot\nabla\times\Bv+\nabla\cdot\Ev=\rhov+\jv\cdot k,
\end{equation}
together with one of the equations $k\times W_1=0$, or $k\times
W_2=0$, that is
\begin{equation}\label{kw1=0}
k\times\left[(k\times\nabla\times\Bv)+\nabla\times\Ev-k\times\jv\right]=0
\quad (k\times W_1=0),
\end{equation}
or
\begin{equation}\label{kw2=0}
k\times\left[\nabla\times\Bv-k\times\nabla\times\Ev-\jv\right]=0\quad
(k\times W_2=0).
\end{equation}
The following proposition concludes our discussion on the
reduction of the Vlasov-Maxwell system to a set of evolution
equations and a set of constraint equations on the initial data.
\begin{Proposition}
The following assertions are equivalent:
\begin{itemize}
\item[(1)]$(\fv,\Ev,\Bv)$ is a solution to the initial value
problem for (\ref{vlasov})--(\ref{constrainteq2})
\item[(2)]$(\fv,\Ev,\Bv)$ is a solution to the initial value
problem for (\ref{vlasov})--(\ref{constrainteq2}) and the initial
data satisfy (\ref{temp1})-(\ref{temp2}) \item[(3)]$(\fv,\Ev,\Bv)$
is a solution to the initial value problem for
(\ref{vlasov})--(\ref{maxwell2}) and the initial data satisfy
(\ref{scalarconst})-(\ref{kw1=0})
 \item[(4)]$(\fv,\Ev,\Bv)$
is a solution to the initial value problem for
(\ref{vlasov})--(\ref{maxwell2}) and the initial data satisfy
(\ref{scalarconst}) and (\ref{kw2=0})
\end{itemize}
\end{Proposition}
\noindent\textit{Proof: }We already proved that
(1)$\Leftrightarrow$(2)$\Rightarrow$(3) and
(3)$\Leftrightarrow$(4). Thus it is sufficient to establish
(4)$\Rightarrow$(1). It is a simple exercise of vector algebra to
show that (\ref{scalarconst}) are satisfied for all times provided
they are satisfied at time $v=0$ and $\Ev,\Bv,\rhov,\jv$ satisfy
(\ref{maxwell1}), (\ref{maxwell2}) and (\ref{localconsingmass}).
The latter holds in virtue of the Vlasov equation (\ref{vlasov}).
Moreover, (\ref{constrainteq1}) follows from (\ref{maxwell1}) and
the second equation in (\ref{scalarconst}), while
(\ref{constrainteq2}) follows from (\ref{maxwell2}) and the first
equation in (\ref{scalarconst}). Thus $(\fv,\Ev,\Bv)$ is a
solution of (\ref{vlasov})--(\ref{constrainteq2}) and since
(\ref{temp1})-(\ref{temp2}) are satisfied at $v=0$, then it is
also a solution of the initial value problem.\prfe

The no-incoming radiation condition in the coordinates $(v,x)$
reads as in the following
\begin{Definition}\label{noincradcon}
A global solution of (\ref{vlasov})--(\ref{sourcedef}) is said to
satisfy the no-incoming radiation condition (NIRC) if, for all
$v_1,\,v_2\geqslant 0$,
\[
\lim_{r\to\infty}\int_{v_1}^{v_2}\int_{|x|=r}k\cdot\left[\Ev\times\Bv\right](v,x)\,dS_r(x)\,dv=0.
\]
A local solution in the interval $[0,V[$, $V>0$, satisfies NIRC if
the above limit is zero for all $v_1,v_2\in [0,V[$.
\end{Definition}
We impose NIRC only in the future, since our purpose is to study
the initial value problem forward in time. Likewise we may
introduce the concept of outgoing radiation as in \cite{C3}.
\begin{Definition}\label{outradiation}
The outgoing radiation $\mathcal{E}_\mathrm{out}(v_1,v_2)$ emitted
by a (global) solution of the Vlasov-Maxwell system in the
interval $[v_1,v_2]$ of the advanced time is given by
\[
\mathcal{E}_\mathrm{out}(v_1,v_2)=\lim_{r\to\infty}\int_{v_1}^{v_2}\int_{|x|=r}k\cdot\left[\Ev\times\Bv\right](v+2r,x)\,dS_r(x)\,dv,
\]
provided the limit exists.
\end{Definition}

The energy identity in the coordinates $(v,x)$ reads
\begin{equation}\label{energyid}
\partial_v(e_\wedge+\pv\cdot
k)=-\nabla\cdot\pv,
\end{equation}
where
\[
e_\wedge=\int\sqrt{1+|p|^2}\fv\,dp+\frac{1}{2}|\Ev|^2+\frac{1}{2}|\Bv|^2,\quad
\pv=\int p\,\fv\, dp+\Ev\times\Bv.
\]
The identity (\ref{energyid}) can be proved either by a direct
calculation using the equations (\ref{vlasov})--(\ref{sourcedef}),
or by a simple change of variables in (\ref{localcon}). Next
define
\[
\mathfrak{m}_\wedge(v,r)=\int_{|x|\leqslant
r}(e_\wedge+\mathfrak{p}_\wedge\cdot k)dx.
\]
By (\ref{energyid}), $\mathfrak{m}_\wedge$ satisfies the equations
\begin{equation}\label{dvm1}
\partial_v\mathfrak{m}_\wedge=-\int_{|x|=r}\mathfrak{p}_\wedge\cdot
k \,dS_r(x),\quad
\partial_v\mathfrak{m}_\wedge+\partial_r\mathfrak{m}_\wedge=\int_{|x|=r}e_\wedge
dS_r(x).
\end{equation}
We define the energy $M_\wedge(v)$ on the past light cone at time
$v$ as
\[
M_\wedge(v)=\lim_{r\to\infty}\mathfrak{m}_\wedge(v,r).
\]
The function $\mathfrak{m}_\wedge(v,\cdot)$ is non-decreasing and
so the above limit exists.
\begin{Lemma}\label{mdecreasing}
$M_\wedge$ is a non-decreasing function:
\[
M_\wedge(v_1)\leqslant M_\wedge(v_2), \quad\forall\,\,
v_1\leqslant v_2.
\]
Moreover if the NIRC is satisfied then $M_\wedge(v)$ is constant
for all $v\geqslant 0$.
\end{Lemma}
\noindent\textit{Proof:} For all $v_1\leqslant v_2$ and
$r>v_2-v_1$ we have, integrating the second equation in
(\ref{dvm1}), $\mathfrak{m}_\wedge(v_2,r)\geqslant
\mathfrak{m}_\wedge(v_1,r+v_1-v_2)$ and letting $r\to\infty$ we
prove that $M_\wedge$ is non-decreasing. To show that $M_\wedge$
is constant in the absence of incoming radiation, we use that, by
the first equation in (\ref{dvm1}),
\begin{eqnarray*}
\mathfrak{m}_\wedge(v_2,r)-\mathfrak{m}_\wedge(v_1,r)&=&
-\int_{v_1}^{v_2}\int_{|x|=r}\int p\cdot k\fv\,dp\,dS_r(x)\,dv\\
&&-\int_{v_1}^{v_2}\int_{|x|=r}k\cdot [\Ev\times\Bv]\,dS_r(x)\,dv.
\end{eqnarray*}
By Lemma \ref{supportf}, the first term in the right hand side
vanishes for $r> 2R_0+v_2$, while the second term tends to zero in
the limit $r\to\infty$ by the NIRC. \prfe

From Lemma \ref{mdecreasing} we obtain the following uniqueness
theorem for the Maxwell equations.
\begin{Lemma}\label{uniqueness}
$(\Ev,\Bv)\equiv 0$ is the unique $C^1$ solution of the
homogeneous system
\begin{equation}\label{homeq}
\partial_v(\Ev-k\times\Bv)-\nabla\times\Bv=0,\quad\partial_v(\Bv+k\times\Ev)+\nabla\times\Ev=0,
\end{equation}
which satisfies the NIRC and the initial condition
$(\Ev,\Bv)(0,x)=0$.
\end{Lemma}
\noindent\textit{Proof: }By Lemma \ref{mdecreasing} we have
\begin{eqnarray*}
0&=&2\left[|\Ev|^2+|\Bv|^2+2(\Ev\times\Bv)\cdot k\right]\\
&=&|\Ev\cdot k|^2+|\Bv\cdot
k|^2+|\Ev-k\times\Bv|^2+|\Bv+k\times\Ev|^2.
\end{eqnarray*}
Hence the solution is a plane wave propagating along the
$k-$direction, i.e., the vectors $(\Ev,\Bv,k)$  form an orthogonal
triad. It follows by (\ref{homeq}) that
$\nabla\times\Ev=\nabla\times\Bv=0$ and so, by
(\ref{scalarconst}), $\nabla\cdot\Ev=\nabla\cdot\Bv=0$.
The claim follows.\prfe

By a standard interpolation argument we obtain
\begin{Lemma}\label{interpolation}
If the initial data are chosen such that $M_\wedge(0)$ is bounded
and the solution satisfies NIRC, then
\[
\|(\rhov+\jv\cdot k)(v)\|_{L^{4/3}(\R^3)}\leqslant
CM_\wedge(0),\quad\forall\,\, v\geqslant 0,
\]
where $C$ is a positive constant which depends only
$\|\fvin\|_\infty$.
\end{Lemma}
\noindent\textit{Proof: }We write
\begin{eqnarray*}
\rhov+\jv\cdot k&=&\int_{|p|\leqslant R}(1+\widehat{p}\cdot
k)\fv\,dp+\int_{|p|>R}(1+\widehat{p}\cdot k)\fv\,dp\\
&\leqslant&\frac{8\pi}{3}\|\fvin\|_\infty R^3+R^{-1}\int
p_0\fv\,dp\leqslant C\left(\int p_0\fv\right)^{3/4}\\
&\leqslant& C(e_\wedge+\mathfrak{p}_\wedge\cdot k)^{3/4},
\end{eqnarray*}
where in the second line we choose
\[
R=\left(\|\fvin\|_\infty^{-1}\int p_0\fv\,dp \right)^{1/4}.
\]
The claim follows.\prfe

We shall now briefly discuss the relation between the the total
energy and the energy on the past light cones. The total energy
(\ref{totalcons}) can be rewritten as
\[
M(t)=\lim_{r\to\infty}\mathfrak{m}(t,r),\quad\mathfrak{m}(t,r)=\int_{|x|\leqslant
r}e(t,x)\,dx=\int_{|x|\leqslant
r}e_\wedge(t+|x|,x)\,dx.
\]
By (\ref{localcon}), $\mathfrak{m}(v,r)$ satisfies the equations
\begin{equation}
\partial_v\mathfrak{m}= -\int_{|x|=r}\mathfrak{p}\cdot
k\,dS_r(x),\label{dvmtotal}
\end{equation}
\begin{equation}
\partial_v\mathfrak{m}\pm\partial_r\mathfrak{m}=\int_{|x|=r}(\pm e-\mathfrak{p}\cdot
k)\,dS_r(x)\label{dvm2}.
\end{equation}
The right hand side of (\ref{dvm2}) is non-negative in the $+$ sign case and non-positive in the
$-$ sign case.
\begin{Lemma}\label{Mconst}
The total energy is constant, i.e.,
\[
M(v_2)=M(v_1),\quad \forall\,\, v_1,v_2\geqslant 0.
\]
\end{Lemma}
\noindent\textit{Proof: }Integrating (\ref{dvm2}) with the plus
sign along the characteristics of $\partial_v+\partial_r$ we
obtain $\mathfrak{m}(v_2,v_2+r)\geqslant\mathfrak{m}(v_1,v_1+r)$
for all $v_1\geqslant v_2$, which implies, in the limit
$r\to\infty$, $M(v_2)\geqslant M(v_1)$. On the other hand,
integrating (\ref{dvm2}) with the minus sign along the
characteristics of $\partial_v-\partial_r$ gives
$\mathfrak{m}(v_2,r-v_2)\leqslant \mathfrak{m}(v_1,r-v_1)$, for
all $v_2\geqslant v_1,\,r>v_2$ and so, letting $r\to\infty$,
$M(v_2)\leqslant M(v_1)$. The claim follows.\prfe

We emphasize that for solutions with data on a past light cone it
is not obvious that $M$ is bounded. If it is bounded, then, by
Lemma \ref{Mconst}, it is conserved.  In the latter case, however,
the total energy and the energy on the past light cones might
be different. To see this consider the equation
\begin{equation}\label{energyidentity}
\mathfrak{m}(v,r)=\mathfrak{m}_\wedge(v,r)-\int_v^{v+r}\int_{|x|=r}\mathfrak{p}_\wedge\cdot
k(v',x)\,dS_r(x)\,dv'.
\end{equation}
The latter is obtained by integrating (\ref{energyid}) in time
from $v$ to $v+|x|$ and proceeding as in the proof of Lemma
\ref{massidentity}. By Lemma \ref{supportf} and
(\ref{energyidentity}) we have
\begin{equation}\label{energyidentity2}
\mathfrak{m}(v,r)=\mathfrak{m}_\wedge(v,r)-\int_v^{v+r}\int_{|x|=r}k\cdot[\Ev^\mathrm{out}\times\Bv^\mathrm{out}](v',x)\,dS_r(x)\,dv',
\end{equation}
for $r> 2R_0+v$, where $\Ev^\mathrm{out},\Bv^\mathrm{out}$ is the
field outside the support of the matter. Hence the answer to the
question whether or not $M_\wedge=M$ depends on the decay of the
solutions of (\ref{homeq}) as $r\to\infty$. As we shall discuss in
Section \ref{sphsymsol}, the equality $M_\wedge=M$ holds for
spherically symmetric solutions, as in this case the magnetic
field vanishes identically. An interesting open question is
whether  $M_\wedge=M$ holds in general in the absence of incoming
radiation.

To conclude this section we study the evolution of the energy on
the future light cones. Let
\begin{eqnarray*}
&&e^{\vee}(v,x)=e_{\wedge}(v+2r,x)=e(v+r,x),\\
&&\mathfrak{p}^{\vee}(v,x)=\mathfrak{p}_{\wedge}(v+2r,x)=\mathfrak{p}(v+r,x),
\end{eqnarray*}
which satisfy the equation
\[
\partial_v (e^\vee-\mathfrak{p}^{\vee}\cdot
k)=-\nabla\cdot\mathfrak{p}^\vee.
\]
Now let
\[
\mathfrak{m}^\vee(v,r)=\int_{|x|\leqslant
r}(e^\vee-\mathfrak{p}^\vee\cdot k)\,dx, \quad
M^\vee(v)=\lim_{r\to\infty}\mathfrak{m}^\vee(v,r)
\]
and note the equations
\begin{equation}\label{eqsm}
\partial_v\mathfrak{m}^\vee=-\int_{|x|=r}\mathfrak{p}^\vee\cdot
k\,dS_r(x),\quad
\partial_v\mathfrak{m}^\vee-
\partial_r\mathfrak{m}^\vee=-\int_{|x|=r}e^\vee\,dS_r(x),
\end{equation}
\begin{equation}\label{energyidentity3}
\mathfrak{m}^\vee(v,r)=\mathfrak{m}_\wedge(v,r)-\int_v^{v+2r}\int_{|x|=r}\mathfrak{p}_\wedge\cdot
k(v',x)\,dS_r(x)\,dv'.
\end{equation}
By (\ref{eqsm})-(\ref{energyidentity3}) and the same argument as
in the proof of Lemma \ref{Nveedecreasing} we obtain
\begin{Lemma}\label{Mveedecreasing}
For all $v_1\leqslant v_2$,
\[
M^\vee(v_2)\leqslant M^\vee(v_1).
\]
Moreover
\[
(i)\quad M^\vee(v_2)=M^\vee(v_1) \textnormal{\textit{ iff }} \lim_{r\to\infty}\int_{v_1}^{v_2}\int_{|x|=r} \mathfrak{p}^\vee\cdot k\,\,dS_r(x)\,dv=0;
\]
\[
(ii)\quad M^\vee(v)=M_\wedge (v) \textnormal{\textit{ iff }} \lim_{r\to\infty}\int_v^{v+2r}\int_{|x|=r}\mathfrak{p}_\wedge\cdot
k\,\,dS_r(x)\,dv'=0.
\]
\end{Lemma}
The remarks on $N^\vee$ following the proof of Lemma
\ref{Nveedecreasing} apply to $M^\vee$ as well. In particular, the
difference $M^\vee(v_1)-M^\vee(v_2)$, when it does not vanish,
measures the energy dissipated by the system to future null
infinity in the interval $[v_1,v_2]$ of the advanced time. By (ii)
of Lemma \ref{Mveedecreasing}, this is the sum of two
contributions: an energy lost in form of outgoing radiation by the
electromagnetic field (as given in Definition \ref{outradiation})
and a kinetic energy carried by the particles, which is given by
the limit
\[
\lim_{r\to\infty}\int_{v_1}^{v_2}\int_{|x|=r}p\cdot k\,\fv(v+2r,x,p)\,dp\,dS_r(x)\,dv.
\]
As in the proof of Lemma \ref{zerolimits}, the latter term vanishes if the momentum support of $\fv$ is
uniformly bounded in $v\in\R$, as in this case no
particles can move to future null infinity. Given this
interpretation, it is natural to identify $M^\vee$ as the analogue
of the Bondi mass in General Relativity, see \cite{B}.

\section{Spherically symmetric solutions}\label{sphsymsol}
\setcounter{equation}{0} In spherical symmetry we have
$\nabla\times\Ev=\nabla\times\Bv=0$ and so, by the first equation in (\ref{scalarconst}),
$\nabla\cdot\Bv=0$. Under the additional boundary condition
$\lim_{r\to\infty}\Bv=0$, this implies that the magnetic field
vanishes identically. Moreover, by the second equation in (\ref{scalarconst}),
\begin{eqnarray}
\Ev(v,x)&=&\frac{k}{r^2}\int_0^r(\rhov+\jv\cdot k)(v,r')r'^2\,dr'\nonumber\\
&=&\frac{1}{4\pi}\int\frac{(x-y)}{|x-y|^3}(\rhov+\jv\cdot k)(v,y)\,dy,\label{ess}
\end{eqnarray}
the second equality being valid in spherical symmetry. By abuse of notation we use the same symbol to denote a
spherically symmetric function in spherical and Cartesian
coordinates.
The Vlasov equation reduces to
\begin{equation}
\partial_v\fv+\frac{p}{p_0}\cdot\nabla_x\fv+\frac{\sqrt{1+|p|^2}}{p_0}\Ev\cdot\nabla_p\fv=0.\label{vlasovss}
\end{equation}
In spherical symmetry the particle density is invariant under proper rotations in phase-space. This allows one to write $\fv=\fv(v,r,w,q)$, where $w=(p\cdot k)\in\R$ and $q=|x\wedge p|^2\geqslant 0$,
see \cite{H}. However the Vlasov equation is more conveniently
studied in the coordinates $(x,p)$. Note also the conservation of angular momentum: along characteristics,
\begin{equation}\label{consangmom}
\frac{d}{ds}|x\times p|^2=0.
\end{equation}
In the spherically symmetric
case we have the following global existence theorem.
\begin{Theorem}\label{global}
Let $0\leqslant \fvin\in
C^1_c(\R^3\times\R^3)$ be spherically symmetric and satisfy
\begin{equation}\label{assumption}
F=\inf\{|x\times p|^2:\,(x,p)\in\textnormal{supp}\fvin\}>0;
\end{equation}
there exists a unique, spherically symmetric $\fv\in C^1([0,\infty[\times\R^3\times\R^3)$ solution  of
(\ref{ess})-(\ref{vlasovss}) such that $\fv(0,x,p)=\fvin(x,p)$.
Moreover, there exists a constant $C>0$, depending only on bounds
on the initial datum, such that
\begin{equation}\label{estP}
\mathcal{P}_\wedge(v)\leqslant C.
\end{equation}
\end{Theorem}
Before giving the proof of Theorem \ref{global}, let us observe
the following
\begin{Corollary}
For the solution of Theorem \ref{global},
\[
N^\vee(v)=N(v)=N_\wedge(v)=N_\wedge(0),
\]
\[
M^\vee(v)=M(v)=M_\wedge(v)=M_\wedge(0).
\]
\end{Corollary}
\noindent\textit{Proof: }The equality of the mass functions
follows from Lemma \ref{zerolimits}. Since spherically symmetric
solutions are isolated from incoming radiation, then $M_\wedge(v)$
is constant by Lemma \ref{mdecreasing}. Setting $B_\wedge^{\rm
out}=0$ and letting $r\to\infty$ in (\ref{energyidentity2}), we
have $M(v)=M_\wedge(v)$. Hence it remains to show that
$M^\vee(v)=M_\wedge(v)$, for all $v\geqslant 0$. For this purpose
we use (\ref{energyidentity3}) with $\Bv=0$, that is
\[
\mathfrak{m}^\vee(v,r)=\mathfrak{m}_\wedge(v,r)-\int_v^{v+2r}\int_{|x|=r}\int_{|p|\leqslant
C}p\cdot k\,\fv(v',x,p)\,dp\,dS_r(x)\,dv';
\]
as in the proof of Lemma \ref{zerolimits}, the integral in the
right hand side of this identity vanishes for $r$ large enough and
letting $r\to\infty$ concludes the proof.\prfe

The proof of Theorem \ref{global} is formally identical to the
proof of global existence for the Cauchy problem with data at time
$t=0$ given in \cite[Theorem II]{GSh} (see \cite{H} for the case
of two different species of particle). We shall sketch it for the
sake of completeness, restricting ourselves to derive the main
estimates which lead to the proof. Note however that the
assumption (\ref{assumption}) is not made in \cite{GSh}. Here the
condition (\ref{assumption}) is used to ensure that the
characteristics are $C^1$ in all the parameters. In fact, due to
the presence of the unit vector $k$, the coefficients of the
Vlasov equation are in general discontinuous at $x=0$. But thanks
to (\ref{assumption}) and the conservation of angular momentum,
the solution is supported away from the axis $|x|=0$ and so it can
be defined in a classical sense in terms of the characteristics.
The assumption (\ref{assumption}) can probably be removed by
passing to a weaker solution concept, but we shall not pursue this
here.

For the proof of Theorem \ref{global}, we denote by $C$ any positive constant which depends only on the
initial datum. Moreover we define
\[
R_{\rm min}(v)=\inf\{|x|:\fv(s,x,p)\neq 0,\,0\leqslant s\leqslant
v,\,p\in\R^3\}.
\]
By the conservation of angular momentum and (\ref{assumption}),
\begin{equation}\label{estR}
R_{\rm min}(v)\geqslant\frac{\sqrt{F}}{\mathcal{P}_\wedge(v)}.
\end{equation}
Hence a bound on the momentum support of $\fv$ implies that the particle density vanishes
in a neighbourhood of the axis $|x|=0$.
Now observe that, by (\ref{lqest}) for $q=1$,
\[
|\Ev(t,x)|\leqslant\frac{N_\wedge}{r^2}.
\]
Moreover, the bound $(\rhov+\jv\cdot k)\leqslant C \mathcal{P}_\wedge(v)^{3}$, Lemma \ref{interpolation} and H\"older's inequality imply
\begin{eqnarray*}
|\Ev(t,x)|&\leqslant&\frac{1}{r^2}\left(\int_0^r(\rhov+\jv\cdot k)^{4/3}r'^2\right)^{1/3}\left(\int_0^r(\rhov+\jv\cdot k)^{5/6}r'^2\right)^{2/3}\\
&\leqslant& \frac{C}{r^2}\|\rhov+\jv\cdot
k\|_{L^{4/3}}^{4/9}\mathcal{P}_\wedge(v)^{5/3}r^2\leqslant
C\mathcal{P}_\wedge(v)^{5/3}.
\end{eqnarray*}
Next define
\[
G(v,r)=-\int_r^\infty\min\left(\frac{N_\wedge}{\lambda^2},C\mathcal{P}_\wedge(v)^{5/3}\right)\,d\lambda,\quad
v,r\geqslant 0.
\]
It follows that $G(v,\cdot)\in C^1$ is increasing and
$|\Ev(v,x)|\leqslant\partial_r G(v,r)$, for all $v\geqslant 0$.
Moreover, and since $\mathcal{P}_\wedge(\cdot)$ is non-decreasing,
$\partial_rG(v_1,r)\leqslant\partial_rG(v_2,r)$, for $v_1
\leqslant v_2$. Splitting the integral at
$R=\sqrt{N_\wedge}(C\mathcal{P}_\wedge(v)^{5/3})^{-1/2}$ one
obtains
\[
G(v,0)=-2\sqrt{N_\wedge}\left(C\mathcal{P}_\wedge(v)^{5/3}\right)^{1/2}
\]
and therefore, for all $r_1,r_2\geqslant 0$,
\[
|G(v,r_1)-G(v,r_2)|\leqslant |G(v,0)|\leqslant
C\mathcal{P}_\wedge(v)^{5/6}.
\]
Next we claim that

\vspace{0.3cm}
\noindent($\sharp$) \textit{there exists at most
one $v_0\in [0,\infty[$ such that $\frac{d}{ds}|X(s)|=0$ and if
such $v_0$ exists then $|X(s)|$ has an absolute minimum at
$s=v_0$.}

\vspace{0.3cm}
\noindent This follows because, along
characteristics,
\[
\frac{d}{ds}|X(s)|=\frac{p\cdot k}{p_0},\quad\frac{d}{ds}(p\cdot
k)=\frac{1}{p_0}\left(\sqrt{1+|p|^2}\,E_\wedge\cdot
k+\frac{|p\times k|^2}{|x|}\right)>0.
\]
Since $p\cdot k$ is increasing and $\frac{d}{ds} |X(s)|>0$ (resp.
$<0$) for $p\cdot k>0$ (resp. $<0$), the claim ($\sharp$) is
proved. Now observe that, along characteristics,
\[
\frac{d}{ds}\sqrt{1+p^2}=\frac{p\cdot k}{p_0}|E_\wedge|.
\]
Hence, for all $v\geqslant 0$ and $s\in [0,v]$ we have, denoting
$\mathcal{I}^+=\{\tau\in [0,s]:\frac{d}{d\tau}|X(\tau)|\geqslant
0\}$,
\begin{eqnarray*}
\sqrt{1+P(s)^2}-\sqrt{1+P(0)^2}&=&\int_0^s|E_\wedge\left(\tau,X(\tau)\right)|\frac{P(\tau)\cdot
K(\tau)}{P_0(\tau)}\,d\tau\\
&=&\int_0^s|E_\wedge\left(\tau,X(\tau)\right)|\frac{d}{d\tau}|X(\tau)|\,d\tau\\
&\leqslant&\int_{\mathcal{I}^+}\partial_rG(\tau,|X(\tau)|)\frac{d}{d\tau}|X(\tau)|\,d\tau\\
&\leqslant&\int_{\mathcal{I}^+}\partial_rG(s,|X(\tau)|)\frac{d}{d\tau}|X(\tau)|\,d\tau\\
&=&\int_{\mathcal{I}^+}\frac{d}{d\tau}\left[G(s,|X(\tau)|)\right]\,d\tau.
\end{eqnarray*}
By virtue of ($\sharp$), either $\mathcal{I}^+=[s_1,s]$, for some
$0< s_1< s$, or $\mathcal{I}^+=[0,s]$, or $\mathcal{I}^+$ is
empty. In the first case we obtain
\begin{eqnarray*}
\sqrt{1+P(s)^2}&\leqslant&
\sqrt{1+P(0)^2}+G(s,|X(s)|)-G(s,|X(s_1)|)\\
&\leqslant& \sqrt{1+P(0)^2}+C\mathcal{P}_\wedge(v)^{5/6}
\end{eqnarray*}
and since this is true for all $0\leqslant s\leqslant v$, then
$\mathcal{P}_\wedge(v)\leqslant C(1+\mathcal{P}_\wedge(v)^{5/6})$,
which implies $\mathcal{P}_\wedge(v)\leqslant C$. The other two
cases lead to the same inequality. The bound on the momentum
support of $\fv$ implies, by (\ref{estR}), that the particle
density is supported away from the axis $|x|=0$. This allows one
to define $\fv$ in terms of the characteristics and derive
$L^\infty$ estimates for its derivatives. A standard iteration
scheme completes the proof of the theorem.

\bigskip
\noindent {\bf Acknowledgments:} The author acknowledges support
by the European HYKE network (contract HPRN-CT-2002-00282) and by
the project ``PDE and Harmonic Analysis'', sponsored by Research
Council of Norway (proj. no. 160192/V30).

\end{document}